\documentclass[letterpaper]{article} 
\usepackage{aaai25}  
\usepackage{times}  
\usepackage{helvet}  
\usepackage{courier}  
\usepackage[hyphens]{url}  
\usepackage{graphicx} 
\usepackage{natbib}  
\usepackage{caption} 
\usepackage{color}

\usepackage{algorithm}
\usepackage{algorithmic}

\usepackage{newfloat}
\usepackage{listings}

\usepackage{amssymb}
\usepackage{amsmath}
\usepackage{multirow}
\usepackage{booktabs} 
\usepackage{pgfplots}
\usepackage{subcaption}
\usepackage[thicklines]{cancel}
\DeclareCaptionStyle{ruled}{labelfont=normalfont,labelsep=colon,strut=off} 
\floatstyle{ruled}
\newfloat{listing}{tb}{lst}{}
\floatname{listing}{Listing}
\title{ConSense: Continually Sensing Human Activity\\ with WiFi via Growing and Picking}

\author {
    Rong Li, 
    Tao Deng\thanks{Tao Deng and Siwei Feng are the corresponding authors.}, 
    Siwei Feng\footnotemark[1], 
    Mingjie Sun, 
    Juncheng Jia
}
\affiliations {
    School of Computer Science and Technology, Soochow University, China\\
    rli23@stu.suda.edu.cn, \{dengtao, swfeng, mjsun, jiajuncheng\}@suda.edu.cn
}

\begin{document}

\maketitle

\begin{abstract}
WiFi-based human activity recognition (HAR) holds significant application potential across various fields. 
To handle dynamic environments where new activities are continuously introduced, WiFi-based HAR systems must adapt by learning new concepts without forgetting previously learned ones.
Furthermore, retaining knowledge from old activities by storing historical exemplar is impractical for WiFi-based HAR due to privacy concerns and limited storage capacity of edge devices.
In this work, we propose ConSense, a lightweight and fast-adapted exemplar-free class incremental learning framework for WiFi-based HAR.
The framework leverages the transformer architecture and involves dynamic model expansion and selective retraining to preserve previously learned knowledge while integrating new information.
Specifically, during incremental sessions, small-scale trainable parameters that are trained specifically on the data of each task are added in the multi-head self-attention layer.
In addition, a selective retraining strategy that dynamically adjusts the weights in multilayer perceptron based on the performance stability of neurons across tasks is used.
Rather than training the entire model, the proposed strategies of dynamic model expansion and selective retraining reduce the overall computational load while balancing stability on previous tasks and plasticity on new tasks.
Evaluation results on three public WiFi datasets demonstrate that ConSense not only outperforms several competitive approaches but also requires fewer parameters, highlighting its practical utility in class-incremental scenarios for HAR.
\end{abstract}

\begin{links}
\link{Code}{https://github.com/kikihub/consense}
\end{links}

\vspace{-0.45cm} %
\section{Introduction}

Human activity recognition (HAR) technologies have broad use in scenarios such as medical monitoring \cite{ge2022contactless}, smart homes \cite{jobanputra2019human}, and security detection \cite{lolla2019wifi}.
Traditional video surveillance faces challenges related to privacy, field of view limitations, and lighting conditions.
In contrast, wireless signal sensing enables non-invasive monitoring to safeguard privacy.
WiFi stands out as an optimal choice for implementing HAR due to its widespread availability and small hardware requirements \cite{guo2019wiar, qian2017widar}.

For WiFi-based HAR, conventional deep learning (DL) models \cite{xia2020lstm, abuhoureyah2024wifi} struggle with identifying novel activities because these systems depend on static models that cannot adjust to emerging human activities.
Thus, there is a pressing need to design DL models that can adeptly adapt to dynamically changing environments. 
Straightforward fine-tuning on DL models with new data without access to the original training data can lead to catastrophic forgetting, where previously learned knowledge is overwritten.

\begin{figure}[!t]
\centering
\includegraphics[width=0.98\columnwidth]{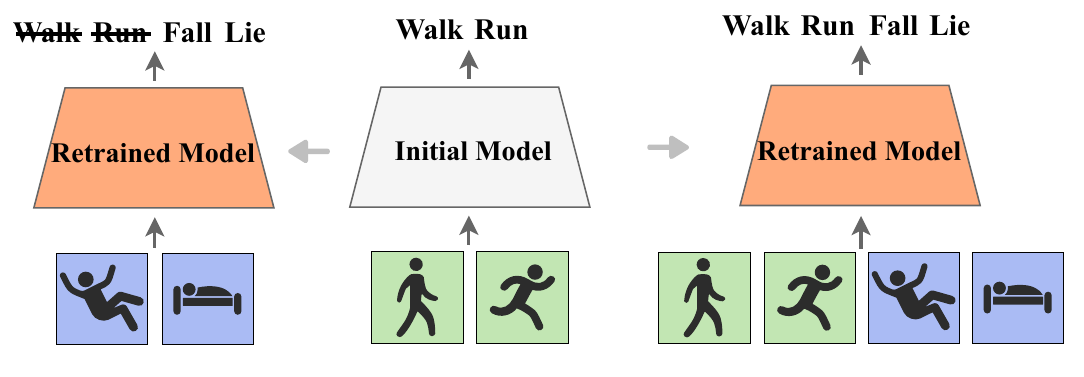}
\vspace{-0.25cm} 
\caption{The initial model in the center is trained on two activities, walk and run. The retrained model on the left, fine-tuned with two new activities, fall and lie, loses the ability to recognize walk and run. In contrast, the retrained model on the right, which incorporates training data from both old and new activities, can recognize all four activities.}
\label{fig_cil}
\vspace{-0.45cm} 
\end{figure}

As illustrated in Figure \ref{fig_cil}, the model retrained on new activities (fall and lie) fails to recognize prior activities (walk and run). An alternative is to retrain a new model using data from both new and previous activities.
However, storing historical data is challenging due to strict data privacy policies \cite{li2014achieving} that limit unregulated storage and transfer, as well as the storage constraints of edge devices \cite{hernandez2020lightweight}.
These challenges require the development of HAR systems that can continually sensing without storing user data, thereby ensuring user privacy and system effectiveness.

Exemplar-free class-incremental learning (EFCIL) \cite{li2017learning,zhu2022self,goswami2024resurrecting} aims to recognize both old and new classes without retaining exemplars from previous classes, addressing concerns regarding privacy and storage.
While extensively explored in computer vision, applying existing EFCIL designed for computer vision tasks to WiFi-based HAR faces distinctive challenges.
Unlike images, WiFi, being a wireless signal, undergoes subtle and time-sensitive changes due to human activities, making stable feature extraction difficult.
This challenge is exacerbated by the continuous and rapidly changing nature of WiFi data, which lacks clear spatial references. Consequently, it's crucial to enhance models that can concurrently capture spatial and temporal characteristics in time series to adapt to the dynamic nature of WiFi data.
Moreover, existing EFCIL approaches often require significant computational resources and lengthy training times, limiting their practicality for resource-constrained edge devices.
This creates an urgent need for lightweight, fast-training continual sensing models that can efficiently manage computation and storage resources for WiFi-based HAR.

To solve these challenges, in this paper we propose ConSense, a continual dynamic adaptive learning framework for WiFi-based HAR.
To capture temporal and spatial relationships in sequential data, ConSense leverages the transformer architecture, which is particularly suitable for processing data with complex spatio-temporal characteristics.
Additionally, ConSense preserves previously learned knowledge while integrating new information by \textbf{growing} with \textbf{dynamic model expansion} and \textbf{picking} with \textbf{selective retraining}.
Specifically, we add small-scale trainable parameters, referred to as prefixes, within the multi-head self-attention (MHSA) layer. These prefixes are custom-designed and trained specifically for data corresponding to each individual task, allowing the model to effectively capture and retain key task-specific knowledge.
This training strategy ensures that the unique features of each task are understood and preserved within the model architecture.
In addition, a selective retraining strategy is employed, which dynamically adjusts the weights of neurons in the multilayer perceptron (MLP) based on their performance across different tasks.
By monitoring the stability of neuron, this module identifies which aspects of information should be maintained over time and which should be adjusted to accommodate new data.
This selective weighting not only helps the model acquire new knowledge without overwriting existing information but also enhances the system's overall adaptability.
These strategies allow ConSense to maintain plasticity and stability during continual sensing while enabling fast training by updating a smaller set of task-specific parameters instead of retraining the entire model.

We validate our proposed framework on three publicly available WiFi datasets, confirming that ConSense exceeds the performance of other models while utilizing fewer parameters.



\vspace{-0.45cm} %
\section{Related Work}


Static DL models for WiFi cannot adapt to the dynamic environment where new activities are constantly introduced.
In computer vision, some works proposed class incremental learning to solve this aspect.
Class-incremental learning (CIL) is classified as three categories: 1) rehearsal-based method, which preserves knowledge by replaying exemplars from past tasks \cite{rebuffi2017icarl, hou2019learning, wu2019large}; 2) regularization-based method, which uses penalties to maintain critical parameters but struggles with lengthy task sequences \cite{kirkpatrick2017overcoming, yang2021cost, saha2021gradient}, and 3) dynamic architecture-based method, which expands the model for new tasks but can be resource-intensive \cite{rusu2016progressive, verma2021efficient, douillard2022dytox}. 
Many existing CIL methods rely on storing exemplars from previous tasks to address catastrophic forgetting. However, the limited storage and computational power of commonly used edge devices, combined with privacy concerns around retaining user data, restrict the applicability of these methods in real-world scenarios.

Some works propose exemplar-free CIL (EFCIL) to avoid the need to retain exemplars.
Li \textit{et al.} \cite{li2017learning} introduced knowledge distillation (KD) for CIL.
However, KD has limited effects when only new data is used.
Gao \textit{et al.} \cite{gao2022r} proposed a new framework that separates representation and classifier learning, thus improving model inversion to synthesize data for previous tasks.
Asadi \textit{et al.} \cite{asadi2023prototype} introduced prototype-sample relation distillation by combining supervised contrastive loss \cite{khosla2020supervised}, self-supervised learning \cite{liu2021self}, and class prototype evolution techniques \cite{de2021continual}. 
By jointly learning representations and class prototypes, they effectively reduce catastrophic forgetting and maintain the relevance as well as the embedding similarity of old class prototypes.
The above methods mainly used ResNet and other convolutional neural network (CNN)-based models.
In general, the two dimensions of WiFi (time stamps and channel state) are fundamentally different from the two dimensions of images that contains spatial information.
Using a two-dimensional kernel to extract spatial patterns from WiFi data would result in poor feature extraction.

Although many CNN and ResNet-based approaches have been developed for EFCIL, transformer-based EFCIL remains a relatively unexplored area.
Roy \textit{et al.} \cite{roy2023exemplar} adapted the transformer's MHSA layers with convolution operations for new tasks. However, their proposed method is unsuitable for WiFi data, as it relies on image augmentation, whereas WiFi augmentation strategies involve temporal delays and frequency shifts, leading to performance degradation.
Zhang \textit{et al.} \cite{zhang2023csi} and Ding \textit{et al.} \cite{ding2023passive} investigated WiFi-based HAR using incremental learning. Zhang \textit{et al.} employed retained exemplars and distillation loss to preserve activity knowledge, while Ding \textit{et al.} introduced an enhancement CNN with attention and dual-loss functions. However, Ding's approach processes only one category at a time, limiting its applicability.
To overcome the limitations of the above methods, we propose a new model specifically designed for WiFi-based HAR, which can more effectively adapt to the dynamic changes in WiFi data while reducing the need for data storage.

\begin{figure*}[t]
\centering
\includegraphics[width=0.86\textwidth]{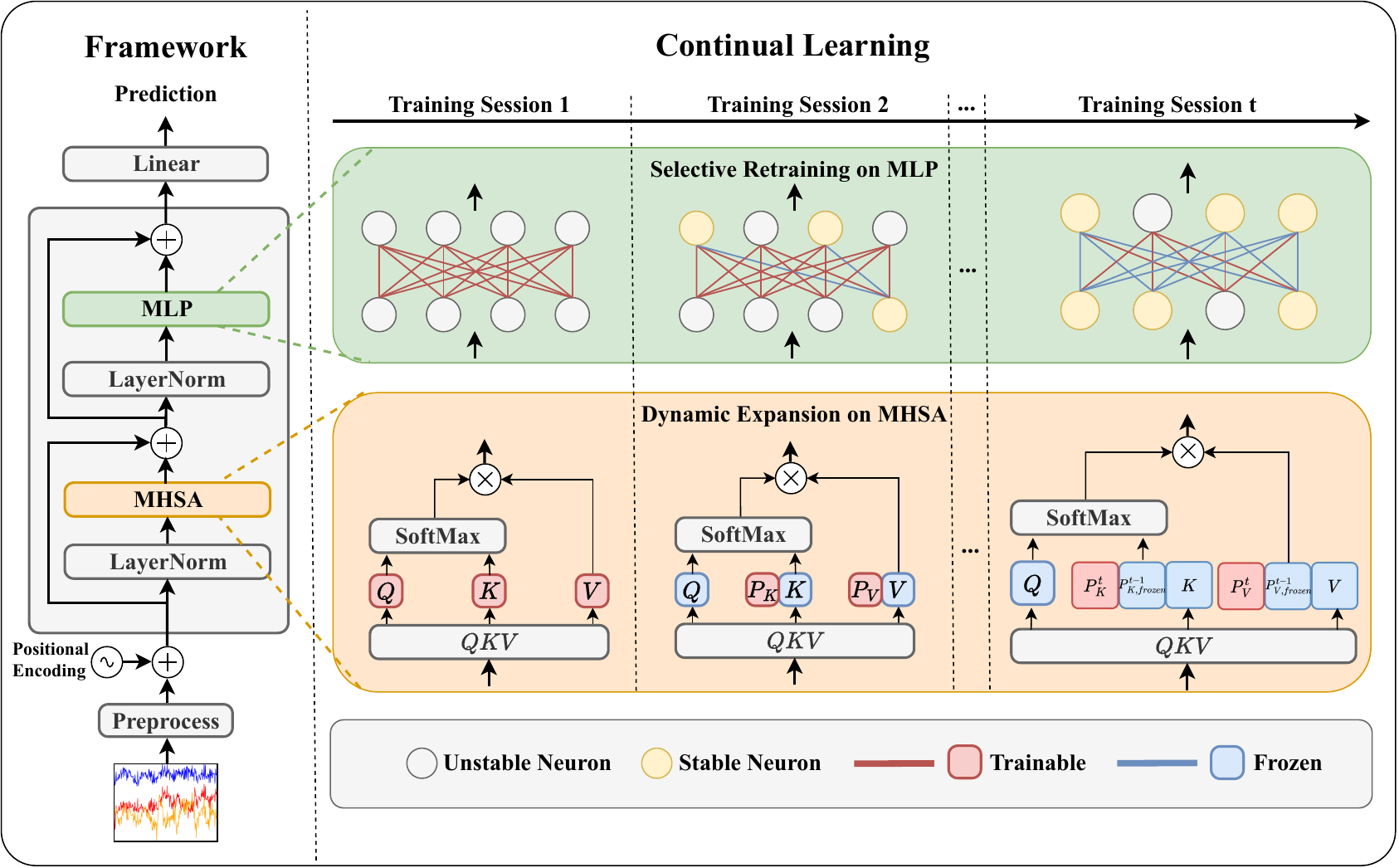} 
\vspace{-0.25cm} 
\caption{Architecture of ConSense. Left part contains the framework. Right part details how the model dynamically expands and selectively retrains during continual learning from training session $1$ to training session $t$.
As new tasks are introduced, the model dynamically expands with new prefixes in the MHSA layer.
In the MLP, a selective retraining strategy is implemented to adjust neuron weights, preserving learned outcomes from stable neurons while updating unstable neurons to accommodate new tasks.}
\label{fig_architecture}
\vspace{-0.45cm} 
\end{figure*}

\vspace{-0.45cm} %
\section{Method}

In EFCIL, a model sequentially learns tasks $\{T_t\}_{t=1}^T$, each introducing a unique set of classes $C_t$, with no class overlap between tasks. The model only accesses the current task's training set $\{X_t, Y_t\}$, where $X_t$ are samples and $Y_t$ are corresponding labels, without storing previous exemplars. The objective of EFCIL is to maximize classification accuracy across all classes encountered up to $T_t$.

The proposed framework performs dynamic model expansion and selective retraining to preserve learned knowledge and learn new classes. During incremental sessions, trainable prefixes are added to the MHSA layer, and a dynamic selective retraining strategy adjusts MLP weights based on neuron performance stability across tasks. The overall framework is illustrated in Figure \ref{fig_architecture}.

\subsection{Preliminaries}

\subsubsection{Input}

WiFi-based HAR utilizes channel state information (CSI) to capture subtle variations in signal characteristics like phase and amplitude, which arise due to environmental interactions and human movement. CSI effectively tracks these changes across multiple subcarrier frequencies, offering a precise method for activity recognition by analyzing how signals interact with physical obstacles and their dynamic alterations in a given space.

\subsubsection{Architecture}

The proposed framework employs a transformer-based architecture, which consists of three main components: positional encoding, a MHSA layer, a MLP.

Sequential information is crucial for recognizing reverse actions such as sitting down and standing up. Traditional absolute or relative position encodings assign a unique and highly distinctive code to each individual point, which can introduce noise. To enhance the capture of sequential information for activities, we employ Gaussian range encoding \cite{li2021two}. This positional encoding method assigns multiple encoding ranges to each position in the data. It allows dynamic adjustments during training based on Gaussian distributions characterized by means $\mu$ and standard deviations $\sigma$. 

MHSA's input sequence, $X \in \mathbb{R}^{n \times d}$, has undergone Gaussian range encoding, where $n$ denotes the temporal dimension and $d$ the spatial dimension of CSI, matching the original input CSI dimensions.
Each attention head $h_i$, where $i$ represents the index of the attention head, uses three matrices $W_i^Q$, $W_i^K$, and $W_i^V$ to transform $X$ into queries $Q_i$, keys $K_i$, and values $V_i$:
\begin{equation}\label{eq_attention}
\text{Attention}(Q_i, K_i, V_i) = \text{softmax}\left(\frac{Q_i K_i^T}{\sqrt{d_k}}\right) V_i,
\end{equation}
where $d_k$ represents the dimension of the key, query and value vectors. $Q_i = X W_i^Q$, $K_i = X W_i^K$, $V_i = X W_i^V$. The final output is generated by concatenating the results from all heads and transforming them through a linear transformation, combining the individual outputs into a comprehensive representation for the entire sequence. The output dimension is the same as the input dimension.

The MLP used in our framework is a type of feedforward neural network, consisting of one input layer, two hidden layers, one output layer, and ReLU activation functions between each layer.

\subsection{Training Procedure}

The training pipeline of ConSense includes an initial training stage and incremental training stages.
During the initial training stage $T_1$, the model learns the first batch of classes $C_1$, with all model parameters including the weights in the MHSA layer, the MLP and the classifier being trainable.
When the initial training stage is completed, the parameters in the MHSA layer are frozen as frozen as $W^{(MHSA)}_{frozen} = \{W^Q_{frozen}, W^K_{frozen}, W^V_{frozen}\}$, representing the frozen weights for query, key and value from the MHSA layer.
In incremental session $t$, we dynamically expand the MHSA layer and selectively retrain the MLP to balance plasticity and stability with minimal computation.
For MHSA, new trainable prefixes $P^t = \{P^t_{K,frozen},P^t_{V,frozen}\}$ are added, working with frozen weights $W^{(MHSA)}_{frozen}$ and previous task prefixes $P^{t-1}_{frozen}$ to adapt to new classes $C_t$. $P^{t-1}_{frozen} = \{P^{t-1}_{K,frozen}, P^{t-1}_{V,frozen}\}$ represent the frozen key and value prefixes from the previous task.
In the MLP, we average neuron activation values, identify stable neurons, and use freeze mask to update only unstable neurons.
After completing task $T_t$, the previous task prefixes $P^{t-1}_{frozen}$ are updated to incorporate the new prefixes from task $T_t$. Specifically, $P^{t-1}_{frozen}$ is updated as follows: $P^t_{frozen} = \{ P^{t}_{K,frozen}, P^{t}_{V,frozen} \}$, where $P^{t}_{K,frozen} = [P^{t}_K, P^{t-1}_{K,frozen}]$ and $P^{t}_{V,frozen} = [P^{t}_V, P^{t-1}_{V,frozen}]$, where [,] denotes the concatenate operation.
Details on dynamic expansion on MHSA and selective retraining on MLP are described as follows.

\subsubsection{Input}
For model input, CSI is structured into dimensions $T_{temp} \times C_{ch}$, where $T_{temp}$ represents the temporal dimension, defined as $T_{temp} = vs$ ($v$ represents the frequency of data packet collection, and $s$ represents the sampling time for an action). The channel dimension $C_{ch}$, defined as $C_{ch} = eg$ ($e$ represents the number of transmitting and receiving antenna combinations, and $g$ represents the number of subcarriers per antenna pair).

\subsubsection{Dynamic Expansion on MHSA}

The goal of acquiring new knowledge while retaining old knowledge without storing exemplars can be achieved by adding task-specific prefixes to the MHSA layers. This method allows knowledge transfer between tasks without significantly changing the model's original parameters. By keeping the model's parameters fixed and updating only the prefixes, the system avoids catastrophic forgetting while maintaining adaptability. Specifically, each MHSA layer has $H$ attention heads, and adding prefixes to these layers enables class-incremental learning for new tasks. Concretely, we have
\begin{align}
{W_i^K}' = [P^t_K, {P^{t-1}_{K,frozen}}, W_{frozen}^K],\notag\\
{W_i^V}' = [P^t_V, {P^{t-1}_{V,frozen}}, W_{frozen}^V].
\end{align}
The output of a head in the self-attention layer is formulated as:
\begin{equation}
\text{head}_i = \text{Attention}(Q_i, {K_i}', {V_i}'), 
\end{equation}
where ${Q_i} = X{W_{frozen}^Q}, {K_i}' = X{W_i^K}', {V_i}' = X{W_i^V}'$, and $\text{Attention}()$ is defined in Eq.\ref{eq_attention}.
To effectively integrate prior knowledge with new information, the model sequentially concatenates the new trainable prefixes $P^t_K$ and $P^t_V$ with the previously frozen prefixes $P^{t-1}_{K,frozen}$ and $P^{t-1}_{V,frozen}$. These concatenated prefixes are then merged with the consistently frozen weights $W_{frozen}^K$ and $W_{frozen}^V$, employing concatenation to ensure a seamless transition and retention of learned features across tasks.

While prefixes are suitable for class-incremental learning tasks, their random initialization can lead to unstable performance due to varying initial weights. To address this, we took inspiration from parallel attention design \cite{yu2022towards}, which uses a parallel adapter to stabilize prefixes. 
Specifically, with input sequence $X$, the prefix generation is formulated as:
\begin{equation}
P_K, P_V = Adapter(X) = Tanh(X W_{down})W_{up}, 
\end{equation}
where $Tanh$ is the activation function, and $W_{down}$ and $W_{up}$ are the parameters of the parallel adapter's scaling layers. $W_{down}$ is a linear transformation layer that reduces the dimensionality of $X$, and $W_{up}$ is another linear transformation that expands the transformed output.

\subsubsection{Selective Retraining on MLP}

While the MHSA layer captures the temporal features of CSI signals through linear transformations, the added MLP layer introduces non-linear transformations to better capture complex features.
To prevent forgetting issues, we utilize a selective retraining strategy based on neuron activation in all MLP layers.
This method involves three main steps: calculating each neuron's average activation value, identifying stable neurons, and generating freeze masks for parameter updates. The process is applied independently to each of the linear layers in the MLP.

First, given a training set, we calculate the average activation value $\bar{a}_p^{(l)}$ for each neuron in the $l$-th layer, defined as:
\begin{equation}
\bar{a}_p^{(l)} = \frac{1}{B} \sum_{q=1}^{B} a_p^{(q,l)}, 
\end{equation}
where $B$ is the size of the training set, $l$ denotes the layer index, and $a_p^{(q,l)}$ is the activation value of the $p$-th neuron in the $l$-th layer for the $q$-th sample.

Next, by comparing the current activation values with those from the previous task, we identify the set of stable neurons $S^{(l)}$ in each layer, which is:
\begin{equation}
S^{(l)} = \{ p \mid \| \bar{a}_p^{(l,t)} - \bar{a}_p^{(l,t-1)} \|_2 \leq \epsilon \}, 
\end{equation}
where $\epsilon$ is a predefined threshold, and $\bar{a}_p^{(l,t)}$ and $\bar{a}_p^{(l,t-1)}$ represent the average activation values for the current and previous tasks in the $l$-th layer, respectively.

Finally, we generate the freeze mask set $M^{(l)} = \{ M_W^{(l)}, M_b^{(l)} \}$ based on the set of stable neurons $S^{(l)}$ for each MLP layer. Specifically, $M_W^{(l)}$ and $M_b^{(l)}$ are mask matrices corresponding to the weight matrix $W^{(l)}$ and bias vector $b^{(l)}$ in the $l$-th layer, respectively, and are initialized with values set to one. For stable neurons in the set $S^{(l)}$, the corresponding values in $M_W^{(l)}$ and $M_b^{(l)}$ are set to zero.

During backpropagation, these masks are applied across all layers by identifying positions where $M_W^{(l)}$ and $M_b^{(l)}$ have a value of zero. At these positions, the corresponding gradients of $W^{(l)}$ and $b^{(l)}$ are set to zero, ensuring that these parameters are not updated. Parameters that are not frozen continue to be updated normally.

In this manner, it reduces forgetting by preserving the weights of stable neurons and prevents the excessive computational load during new task learning.

\vspace{-0.25cm} %
\section{Experiments}

\subsection{Datasets and Settings}

Datasets with a limited number of categories are not suitable for evaluating class-incremental learning. Therefore, we selected the WiAR \cite{guo2019wiar}, MMFi \cite{yang2024mm}, and XRF \cite{wang2024xrf55} datasets, which offer a broader range of categories. The statistics of these datasets are summarized in Table \ref{tab_dataset}.

\begin{table}[h]
\small
\centering
\begin{tabular}{c c c c c}
\toprule
Dataset & Class & Size & Train & Test \\
\midrule
WiAR & 16 & 270 × 90 & 384 & 96\\
MMFi & 27 & 10 × 342 & 2160 & 540\\
XRF  & 48 & 50 × 270 & 672 & 288\\
\bottomrule
\end{tabular}
\caption{Statistics of the evaluation datasets.  The size of each dataset is denoted as $T_{temp} \times C_{ch}$, where $T_{temp}$ represents the temporal dimension and $C_{ch}$ represents the channel dimension.}
\label{tab_dataset}
\vspace{-0.35cm} 
\end{table}

\noindent \textbf{WiAR} consists of 480 CSI samples, evenly distributed across 16 distinct classes.
We divided the dataset into training and testing subsets at a 4:1 ratio.
After our processing, the sample size is 270 x 90.
Additionally, we organized WiAR into two task types: short task and long task. The short task set includes 5 tasks: the first task covers 8 classes, while the following 4 tasks cover 2 classes each. In contrast, the long task set comprises 8 tasks, with each task consistently including 2 classes.

\noindent \textbf{MMFi} comprises 2700 CSI samples, evenly distributed across 27 classes.
After our processing, the sample size is 10 x 342.
In MMFi, the short task category includes a total of 6 tasks. The first task covers 12 classes, while each of the next five tasks covers 3 classes. In contrast, the long task category consists of 9 tasks, with each task handling 3 classes.

\noindent \textbf{XRF} initially includes 55 classes, with 7 dedicated to dual-person actions. We exclude these dual-person classes due to our focus on single-person activities, leaving 48 classes and 960 CSI samples.
Each class contains 20 samples, with 14 samples per class allocated for training and the remaining 6 used for testing.
After our processing, the sample size is 50 x 270.
In XRF, the short task category consists of 5 tasks: the first task covers 24 classes, while each of the subsequent 4 tasks contains 6 classes. In contrast, the long task category is organized into 8 tasks, each responsible for analyzing 6 classes.

\subsection{Baselines}

We compare ConSense with five existing EFCIL methods:
(1) LWF \cite{li2017learning} uses knowledge distillation to mitigate forgetting.
(2) PASS \cite{zhu2021prototype} combines prototype augmentation with self-supervised learning to enhance memory of old classes.
(3) R-DFCIL \cite{gao2022r} synthesizes data for previous classes using model inversion and applies relation-guided representation learning to minimize the domain gap between synthetic and real data.
(4) PRD \cite{asadi2023prototype} introduces a new distillation loss to maintain the relevance of class prototypes during new task learning.
(5) ConTraCon \cite{roy2023exemplar} modifies MHSA layer weights via convolutional operations to adapt the transformer architecture for new tasks.

\subsection{Evaluation Metrics}

We use two metrics, i.e., the average accuracy and average forgetting measure the performance of ConSense on all the classes seen so far. The accuracy after each task, denoted by $A_t$, represents the accuracy over all classes learned up to and including the $t$-th task. Subsequently, the average accuracy across all tasks, represented by $\bar{A}$, is expressed as: 
$\bar{A} = \frac{1}{N} \sum_{t=1}^{N} A_t$,
where $N$ represents the number of tasks.
The average forgetting measure \cite{chaudhry2018riemannian} is used to estimate the forgetting of previous tasks.
For each task $t$, the forgetting measure of predicting previous task $k$ is denoted by $f_k^t$, which is expressed as
$f_k^t = \max_{z \in \{1, \ldots, k-1\}} (\alpha_{z,t} - \alpha_{z,t})$,
where $\alpha_{m,j}$ represents the accuracy of task $j$ after training task $m$.  The average forgetting measure represents the forgetting measure of the last task, denoted by $\bar{F}$, which is expressed as
${\bar{F}} = \frac{1}{N-1} \sum_{k=1}^{N-1} f_k^N$.

\begin{table*}[h]
\small
\centering
\begin{tabular}{c c c c c c c c c c c c c}
\toprule
\multirow{2}{*}{Method} & \multirow{2}{*}{Replay Data} & \multicolumn{3}{c}{WiAR} & \multicolumn{3}{c}{MMFi} & \multicolumn{3}{c}{XRF} \\
\cmidrule(r){3-5} \cmidrule(r){6-8} \cmidrule(r){9-11}
& & Params & $N=5$ & $N=8$ & Params & $N=6$ & $N=9$ & Params & $N=5$ & $N=8$ \\
\midrule
                LWF        & -          & 18.52M & 40.77 & 37.26 & 18.52M & 33.81 & 30.54 & 18.52M & 33.30 & 29.77\\
                PASS       & -          & 11.32M & 59.15 & 40.96 & 11.32M & 45.19 & 39.29 & 11.32M & 48.65 & 35.93\\
                R-DFCIL    & Synthetic  & 12.81M & 60.63 & 57.76 & 12.81M & 50.95 & 47.83 & 12.81M & 49.81 & 44.30\\
                PRD        & -          & 11.75M & 64.58 & 60.23 & 11.75M & 54.46 & 52.22 & 11.75M & 54.44 & 51.57\\
                ConTraCon  & -          & 3.60M  & -     & 48.58 & 2.50M  & -     & 44.08 & 2.10M  & -     & 41.03\\
\textbf{ConSense}   & -          & \textbf{3.35M} & \textbf{91.66} & \textbf{89.85} & \textbf{1.92M} & \textbf{84.42} & \textbf{71.97} & \textbf{1.50M} & \textbf{66.19} & \textbf{65.79}\\
\bottomrule
\end{tabular} 
\caption{The average accuracy $\bar{A}$ (\%) comparison of Our ConSense with other five methods on WiAR, MMFi, and XRF with short task ($N=5$ or $N=6$) and long task ($N=8$ or $N=9$). Params refers to the initial number of parameters of a model, measured in millions. None of the methods utilize real historical data for replay. R-DFCIL employs synthetic data to simulate the replay data.}
\label{tab_accuracy}
\vspace{-0.35cm} 
\end{table*}

\subsection{Implementation Details }
We set the number of Gaussian distributions in the positional encoding to 10.
The values of $\mu$s are uniformly distributed across the temporal dimension for various datasets as follows.
For WiAR, they range from 13.5 to 256.5 with a step size of 27.
For MMFi, they range from 0.5 to 9.5 with a step size of 1.
For XRF, they range from 2.5 to 47.5 with a step size of 5.
The standard deviation of the Gaussian distributions on all the datasets is uniformly set to 8.
The number of stacks in the module is set to 1.
The input dimensions for the three datasets are set to 90, 342, and 270, respectively, while 
maintaining a consistent number of heads at 9 for each, and employing a dropout rate of 0.1.

\begin{table}[h]
\small
\centering
\setlength{\tabcolsep}{3pt} 
\begin{tabular}{c c c c c c c}
\toprule
\multirow{2}{*}{Method} & \multicolumn{2}{c}{WiAR} & \multicolumn{2}{c}{MMFi} & \multicolumn{2}{c}{XRF} \\
\cmidrule(r){2-3} \cmidrule(r){4-5} \cmidrule(r){6-7}
& $N=5$ & $N=8$ & $N=6$ & $N=9$ & $N=5$ & $N=8$ \\
\midrule
                LWF           & 31.46 & 31.38 & 31.47 & 33.07 & 32.91 & 29.98\\
                PASS          & 24.51 & 28.94 & 22.64 & 25.68 & 20.49 & 28.74\\
                R-DFCIL       & 22.84 & 24.83 & 20.27 & 24.74 & 21.05 & 27.63\\
                PRD           & 20.30 & 21.69 & 19.30 & 19.34 & 24.34 & 20.90\\
                ConTraCon     & -     & 28.15 & -     & 29.88 & -     & 25.59\\
\textbf{ConSense}      & \textbf{14.31} & \textbf{12.89} & \textbf{16.28} & \textbf{17.99} & \textbf{19.51} & \textbf{18.09}\\
\bottomrule
\end{tabular}
\caption{The average forgetting measure $\bar{F}$ (\%) comparison of Our ConSense with other five methods on WiAR, MMFi, and XRF with short and long tasks (lower is better).}
\label{tab_forgetting}
\vspace{-0.35cm} 
\end{table}

Our method is implemented by PyTorch \cite{paszke2019pytorch} and trained on NVIDIA A5000 GPU with 32GB memory.
The optimizer chosen is Adam \cite{kingma2014adam}, with an initial learning rate of 0.001 and a batch size of 16. The model's training cycle is set to 50 epochs.

\subsection{Comparative Results}
\subsubsection{Performance Comparison}

Tables \ref{tab_accuracy} and \ref{tab_forgetting} present the results of the average accuracy $\bar{A}$ and the average forgetting $\bar{F}$, respectively. 
From the two tables, we observe that ConSense significantly outperforms other methods on all the datasets. 
Specifically, in the long task sequences of WiAR dataset, the average accuracy of ConSense surpasses that of other methods by nearly 30\%.
In the short task sequences of MMFi dataset, the average accuracy improvement exceeds 30\%.
Especially compared to LWF, the advantage of ConSense reaches 50\%.
In addition, ConSense achieves a forgetting rate of less than 20\% for both short and long tasks on the three datasets, and outperforms other methods. 
The two insights manifest that ConSense effectively balances plasticity and stability.
The reason is that 
in ConSense, we utilize MHSA and positional encoding. This design particularly adapts to the characteristics of time-series data, such as the patterns and intensity of signal changes in CSI. 
These features pose challenges to traditional image-based network architectures, like Resnet, which primarily optimizes for spatial feature extraction and struggles with the dynamic characteristics of time-series data.
However, for other methods, the knowledge distillation approach of LWF does not fare well in dynamically changing environments, and the synthetic data approach of R-DFCIL fails to accurately capture the true characteristics of CSI.
PRD attempts to mitigate forgetting by maintaining relationships between class prototypes, but the high dynamism and complexity of CSI may render this prototype-based method ineffective.
ConTraCon uses the Transformer architecture to adapt to new tasks, but its success depends on the effectiveness of its attention mechanism.
If this mechanism fails to capture the temporal and frequency domain characteristics of CSI, the results may fall short.
Moreover, its entropy-based task prediction, which relies on image enhancement techniques, is unsuitable for CSI, as CSI variations like temporal delays and frequency shifts don't correspond to visual changes.

\pgfplotscreateplotcyclelist{colorlist1}{%
    {blue, mark=*},
    {brown, mark=*},
    {green, mark=*},
    {orange, mark=*},
    {red, mark=triangle*},
}
\pgfplotscreateplotcyclelist{colorlist2}{%
    {blue, mark=*},
    {brown, mark=*},
    {green, mark=*},
    {orange, mark=*},
    {purple, mark=*},
    {red, mark=triangle*},
}
\begin{figure*}[t]
\centering
\includegraphics[width=0.86\textwidth]{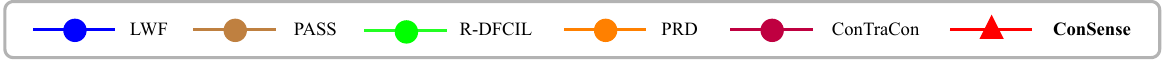} 
\label{fig_legend}
\vspace{-0.35cm} 
\end{figure*}
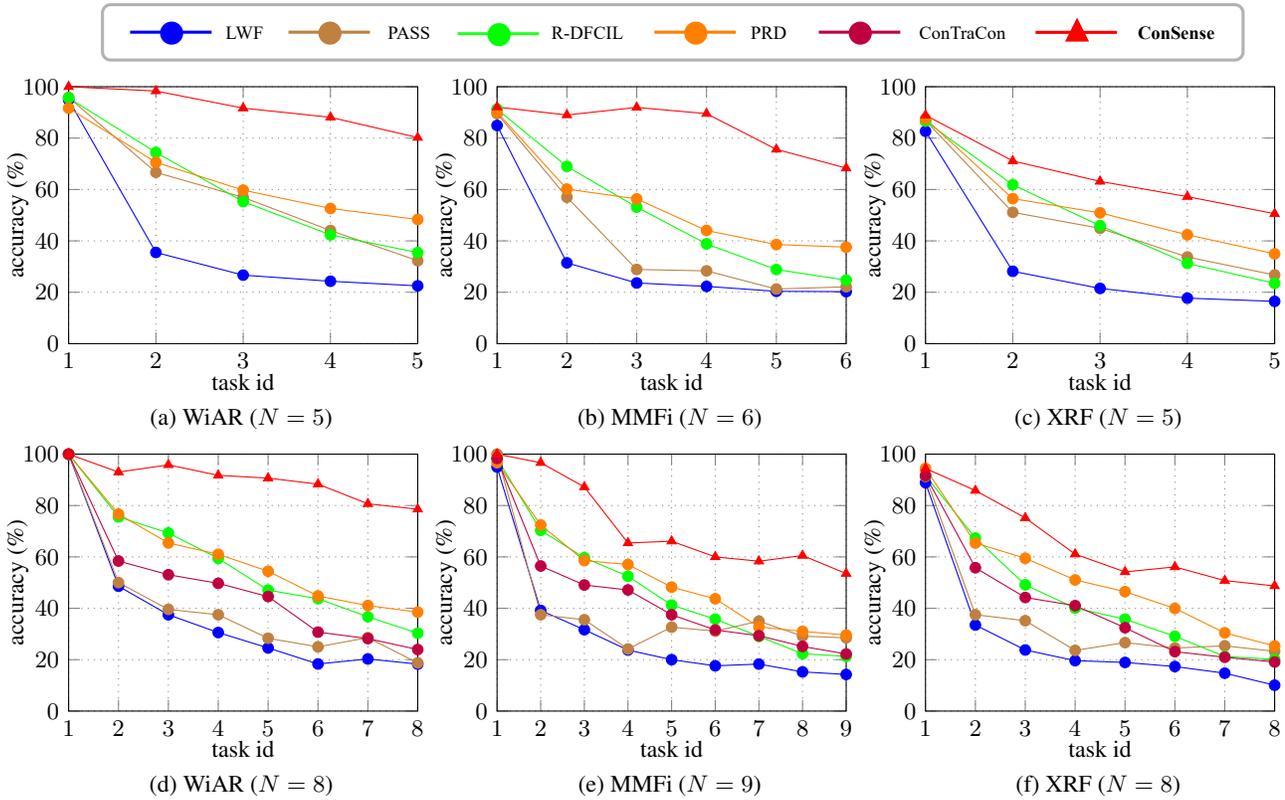
\begin{figure*}[!htb]
    \begin{subfigure}[b]{0.35\textwidth}
        \begin{tikzpicture}
        \begin{axis}[
            width=\textwidth,
            height=5cm,
            xlabel={task id},
            ylabel={accuracy (\%)},
            cycle list name=colorlist1,
            label style={font=\small, inner sep=0pt},
            tick label style={font=\small},
            grid=major,
            grid style={dotted, gray},
            legend pos=south west,
            legend style={fill=none, font=\fontsize{4}{6}\selectfont,legend columns=3, draw=none},
            xlabel style={font=\small, at={(axis description cs:0.5,0.04)}},
            ylabel style={font=\small, at={(axis description cs:0.12,0.5)}},
            xmin=1, ymin=0, 
            xmax=5, ymax=100,
            xtick={1, 2, 3, 4, 5}, 
            ytick={0, 20, 40, 60, 80, 100}, 
        ]
        \addplot coordinates {(1,94.92) (2,35.50) (3,26.67) (4,24.29) (5,22.50)};
        \addplot coordinates {(1,95.83) (2,66.67) (3,56.94) (4,44.04) (5,32.29)};
        \addplot coordinates {(1,95.65) (2,74.43) (3,55.29) (4,42.34) (5,35.46)};
        \addplot coordinates {(1,91.67) (2,70.56) (3,59.74) (4,52.61) (5,48.34)};
        \addplot coordinates {(1,100.00) (2,98.33) (3,91.67) (4,88.10) (5,80.21)};
        \end{axis}
        \end{tikzpicture}
        \caption{WiAR ($N=5$)}
    \end{subfigure}
    \hspace{-2em} 
    \begin{subfigure}[b]{0.35\textwidth}
        \begin{tikzpicture}
        \begin{axis}[
            width=\textwidth,
            height=5cm,
            xlabel={task id},
            ylabel={accuracy (\%)},
            cycle list name=colorlist1,
            label style={font=\small, inner sep=0pt},
            tick label style={font=\small},
            grid=major,
            grid style={dotted, gray},
            legend pos=south west,
            legend style={fill=none, font=\fontsize{4}{10}\selectfont,legend columns=3, draw=none},
            xlabel style={font=\small, at={(axis description cs:0.5,0.04)}},
            ylabel style={font=\small, at={(axis description cs:0.12,0.5)}},
            xmin=1, ymin=0, 
            xmax=6, ymax=100,
            xtick={1, 2, 3, 4, 5,6}, 
            ytick={0, 20, 40, 60, 80, 100}, 
        ]
        \addplot coordinates {(1,84.92) (2,31.45) (3,23.61) (4,22.29) (5,20.37) (6,20.24)};
        \addplot coordinates {(1,89.58) (2,57.00) (3,28.88) (4,28.33) (5,21.25) (6,22.14)};
        \addplot coordinates {(1,91.25) (2,68.98) (3,53.14) (4,38.83) (5,28.86) (6,24.74)};
        \addplot coordinates {(1,90.08) (2,60.15) (3,56.34) (4,44.04) (5,38.58) (6,37.57)};
        \addplot coordinates {(1,92.08) (2,89.00) (3,91.94) (4,89.52) (5,75.62) (6,68.33)};
        \end{axis}
        \end{tikzpicture}
        \caption{MMFi ($N=6$)}
    \end{subfigure}
    \hspace{-2em} 
    \begin{subfigure}[b]{0.35\textwidth}
        \begin{tikzpicture}
        \begin{axis}[
            width=\textwidth,
            height=5cm,
            xlabel={task id},
            ylabel={accuracy (\%)},
            cycle list name=colorlist1,
            label style={font=\small, inner sep=0pt},
            tick label style={font=\small},
            grid=major,
            grid style={dotted, gray},
            legend pos=south west,
            legend style={fill=none, font=\fontsize{4}{10}\selectfont,legend columns=3, draw=none},
            xlabel style={font=\small, at={(axis description cs:0.5,0.04)}},
            ylabel style={font=\small, at={(axis description cs:0.12,0.5)}},
            xmin=1, ymin=0, 
            xmax=5, ymax=100,
            xtick={1, 2, 3, 4, 5}, 
            ytick={0, 20, 40, 60, 80, 100}, 
        ]
        \addplot coordinates {(1,82.66) (2,28.17) (3,21.50) (4,17.71) (5,16.46)};
        \addplot coordinates {(1,86.80) (2,51.11) (3,44.90) (4,33.73) (5,26.73)};
        \addplot coordinates {(1,86.59) (2,61.84) (3,45.91) (4,31.25) (5,23.49)};
        \addplot coordinates {(1,87.51) (2,56.45) (3,50.91) (4,42.38) (5,34.97)};
        \addplot coordinates {(1,88.89) (2,71.11) (3,63.15) (4,57.22) (5,50.58)};
        \end{axis}
        \end{tikzpicture}
        \caption{XRF ($N=5$)}
    \end{subfigure}
    \vskip\baselineskip
    \vspace{-0.35cm} 
    \begin{subfigure}[b]{0.35\textwidth}
        \begin{tikzpicture}
        \begin{axis}[
            width=\textwidth,
            height=5cm,
            xlabel={task id},
            ylabel={accuracy (\%)},
            cycle list name=colorlist2,
            label style={font=\small, inner sep=0pt},
            tick label style={font=\small},
            grid=major,
            grid style={dotted, gray},
            legend pos=south west,
            legend style={fill=none, font=\fontsize{6}{10}\selectfont,legend columns=3, draw=none,at={(-0.01, -0.03)}},
            xlabel style={font=\small, at={(axis description cs:0.5,0.04)}},
            ylabel style={font=\small, at={(axis description cs:0.12,0.5)}},
            xmin=1, ymin=0, 
            xmax=8, ymax=100,
            xtick={1, 2, 3, 4, 5,6,7,8}, 
            ytick={0, 20, 40, 60, 80, 100}, 
        ]
        \addplot coordinates {(1,100.00) (2,48.59) (3,37.50) (4,30.56) (5,24.58) (6,18.33) (7,20.28) (8,18.29)};
        \addplot coordinates {(1,100.00) (2,50.00) (3,39.58) (4,37.50) (5,28.33) (6,25.00) (7,28.57) (8,18.75)};
        \addplot coordinates {(1,100.00) (2,75.56) (3,69.31) (4,59.43) (5,47.09) (6,43.74) (7,36.70) (8,30.31)};
        \addplot coordinates {(1,100.00) (2,76.65) (3,65.43) (4,61.02) (5,54.42) (6,44.79) (7,41.07) (8,38.51)};
        \addplot coordinates {(1,100.00) (2,58.40) (3,53.07) (4,49.72) (5,44.57) (6,30.73) (7,28.21) (8,23.95)};
        \addplot coordinates {(1,100.00) (2,93.02) (3,95.83) (4,91.75) (5,90.67) (6,88.33) (7,80.67) (8,78.58)};
        \end{axis}
        \end{tikzpicture}
        \caption{WiAR ($N=8$)}
    \end{subfigure}
    \hspace{-2em} 
    \begin{subfigure}[b]{0.35\textwidth}
        \begin{tikzpicture}
        \begin{axis}[
            width=\textwidth,
            height=5cm,
            xlabel={task id},
            ylabel={accuracy (\%)},
            cycle list name=colorlist2,
            label style={font=\small, inner sep=0pt},
            tick label style={font=\small},
            grid=major,
            grid style={dotted, gray},
            legend pos=north east,
            legend style={fill=none, font=\fontsize{5}{10}\selectfont,legend columns=2, draw=gray!40,at={(0.96, 0.95)}},
            xlabel style={font=\small, at={(axis description cs:0.5,0.04)}},
            ylabel style={font=\small, at={(axis description cs:0.12,0.5)}},
            xmin=1, ymin=0, 
            xmax=9, ymax=100,
            xtick={1, 2, 3, 4, 5,6,7,8,9}, 
            ytick={0, 20, 40, 60, 80, 100}, 
        ]
        \addplot coordinates {(1,95.00) (2,39.17) (3,31.67) (4,23.75) (5,20.00) (6,17.61) (7,18.29) (8,15.25) (9,14.25)};
        \addplot coordinates {(1,100.00) (2,37.50) (3,35.55) (4,24.16) (5,32.66) (6,31.11) (7,35.00) (8,29.16) (9,28.51)};
        \addplot coordinates {(1,98.33) (2,70.34) (3,59.67) (4,52.48) (5,41.24) (6,35.79) (7,29.07) (8,22.34) (9,21.28)};
        \addplot coordinates {(1,96.66) (2,72.43) (3,58.57) (4,57.05) (5,48.24) (6,43.70) (7,32.75) (8,31.02) (9,29.61)};
        \addplot coordinates {(1,98.33) (2,56.46) (3,49.07) (4,47.15) (5,37.49) (6,31.57) (7,29.33) (8,25.16) (9,22.21)};
        \addplot coordinates {(1,100.00) (2,96.67) (3,87.22) (4,65.42) (5,66.11) (6,60.00) (7,58.33) (8,60.54) (9,53.52)};
        \end{axis}
        \end{tikzpicture}
        \caption{MMFi ($N=9$)}
    \end{subfigure}
    \hspace{-2em} 
    \begin{subfigure}[b]{0.35\textwidth}
        \begin{tikzpicture}
        \begin{axis}[
            width=\textwidth,
            height=5cm,
            xlabel={task id},
            ylabel={accuracy (\%)},
            cycle list name=colorlist2,
            label style={font=\small, inner sep=0pt},
            tick label style={font=\small},
            grid=major,
            grid style={dotted, gray},
            legend pos=north east,
            legend style={fill=none, font=\fontsize{5}{10}\selectfont,legend columns=2, draw=gray!40,at={(0.96, 0.95)}},
            xlabel style={font=\small, at={(axis description cs:0.5,0.04)}},
            ylabel style={font=\small, at={(axis description cs:0.12,0.5)}},
            xmin=1, ymin=0, 
            xmax=8, ymax=100,
            xtick={1, 2, 3, 4, 5,6,7,8}, 
            ytick={0, 20, 40, 60, 80, 100}, 
        ]
        \addplot coordinates {(1,88.89) (2,33.50) (3,23.78) (4,19.63) (5,18.92) (6,17.33) (7,14.74) (8,10.07)};
        \addplot coordinates {(1,91.32) (2,37.50) (3,35.18) (4,23.61) (5,26.66) (6,24.53) (7,25.39) (8,23.26)};
        \addplot coordinates {(1,91.66) (2,67.33) (3,49.14) (4,40.07) (5,35.78) (6,29.10) (7,21.24) (8,20.13)};
        \addplot coordinates {(1,94.44) (2,65.40) (3,59.47) (4,51.04) (5,46.47) (6,39.95) (7,30.46) (8,25.34)};
        \addplot coordinates {(1,91.66) (2,55.78) (3,44.21) (4,41.04) (5,32.45) (6,23.12) (7,20.95) (8,19.09)};
        \addplot coordinates {(1,94.44) (2,85.83) (3,75.19) (4,61.08) (5,54.17) (6,56.11) (7,50.79) (8,48.71)};
        \end{axis}
        \end{tikzpicture}
        \caption{XRF ($N=8$)}
    \end{subfigure}
    \caption{Accuracy comparison of ConSense with other five methods on each task. The x-axis represents the $t$-th task, and the y-axis represents the accuracy of the $t$-th task, i.e., $A_t$.}
    \label{fig_acc}
    \vspace{-0.32cm} 
\end{figure*}

\subsubsection{Parameters Comparison }
Moreover, ConSense exhibits a marked reduction in model parameters compared to other methods. 
Specifically, on the WiAR dataset, the parameter count of ConSense is comparable to ConTraCon, but at least three times less than other methods. 
Notably, on the MMFi and XRF datasets, the parameter count of ConSense is even only one-sixth of that of other methods excluding ConTraCon.
This advantage manifests that ConSense is especially suitable for the deployment of edge devices, e.g., WiFi-based HAR terminal.

\subsubsection{Accuracy of Each Task }
In Figure \ref{fig_acc}, we compare the accuracy of ConSense with other methods on each task on three datasets. 
We observe that the accuracy of the initial task for all the methods is comparable, and the accuracy of ConSense 
significantly outperforms other methods in subsequent tasks. 
This demonstrates that our method achieves a better balance between knowledge forgetting and acquisition when dealing with CSI. 
In addition, the performance gain of ConSense compared to other methods widens with the increase of the number of tasks. 
For example, in Figure \ref{fig_acc}(e), 
at the fifth task, the gain of ConSense compared to PRD is approximately 10\%, while at the ninth task, the gain has reached 20\%.
This trend emphasizes the effectiveness of ConSense in handling extended task sequences, highlighting its robustness in continually sensing.

\subsection{Ablation Test }

\begin{table}[h]
\small
\centering
\begin{tabular}{c c c c c}
\toprule
Dataset & Strategy 1 & Strategy 2 & $A_T$ & $\bar{A}$ \\
\midrule
\multirow{4}{*}{WiAR} & $\times$ & $\times$ & 36.45 & 52.08 \\
                      & $\checkmark$ & $\times$ & 51.03 & 67.70 \\
                      & $\times$ & $\checkmark$ & 44.79 & 59.37 \\
                      & $\checkmark$ & $\checkmark$ & \textbf{78.58} & \textbf{89.85} \\
\midrule
\multirow{4}{*}{MMFi} & $\times$ & $\times$ & 25.91 & 46.27 \\
                      & $\checkmark$ & $\times$ & 38.87 & 60.15 \\
                      & $\times$ & $\checkmark$ & 34.24 & 54.78 \\
                      &  $\checkmark$ & $\checkmark$ & \textbf{53.52} & \textbf{71.97} \\
\midrule
\multirow{4}{*}{XRF} & $\times$ & $\times$ & 22.56 & 40.96 \\
                      & $\checkmark$ & $\times$ & 41.66 & 53.12 \\
                      & $\times$ & $\checkmark$ & 30.90 & 48.60 \\
                      &  $\checkmark$ & $\checkmark$ & \textbf{48.71} & \textbf{65.79} \\
\bottomrule
\end{tabular}
\caption{Ablation study of two strategies for long task on three datasets. Strategy 1 represents dynamic expansion on MHSA. Strategy 2 represents selective retraining on MLP.}
\label{ablation}
\vspace{-0.4cm} 
\end{table}

\subsubsection{Effect of Dynamic Expansion and Selective Retraining}

Table \ref{ablation} shows that the two strategies, dynamic expansion on MHSA and selective retraining on MLP, significantly enhance the performance of ConSense in long-term EFCIL ablation experiments.
More specifically, dynamic expansion significantly enhances the accuracy of the last task and average task accuracy over all the datasets.
It achieves this by adding trainable prefixes to the multi-head self-attention layers, thereby emphasizing its robustness against forgetting and adaptability to new tasks.
Compared to dynamic expansion, while the performance gains observed with selective retraining are less pronounced, it still contributes to model performance enhancement, particularly through stabilizing trained parameters in specific scenarios. 
\vspace{-0.3cm} 

\pgfplotscreateplotcyclelist{colorlist3}{%
    {blue, mark=*},
    {green, mark=*},
    {red, mark=triangle*},
}

\begin{figure}[h]
\centering
\setlength{\abovecaptionskip}{0.2cm}
\begin{tikzpicture}
\begin{axis}[
    width=0.4\textwidth,
    height=5cm,
    xlabel={task id},
    ylabel={accuracy (\%)},
    cycle list name=colorlist3,
    label style={font=\small, inner sep=0pt},
    tick label style={font=\small},
    grid=major,
    grid style={dotted, gray},
    legend pos=south west,
    legend style={fill=none, font=\fontsize{7}{10}\selectfont,legend columns=3, draw=gray!40,at={(0.04, 0.2)}},
    xlabel style={font=\small, at={(axis description cs:0.5,0.04)}},
    ylabel style={font=\small, at={(axis description cs:0.08,0.5)}},
    xmin=1, ymin=0, 
    xmax=9, ymax=100,
    xtick={1, 2, 3, 4, 5,6,7,8,9}, 
    ytick={0, 20, 40, 60, 80, 100}, 
]
\addplot coordinates {(1,100.0) (2,88.33) (3,75.78) (4,62.92) (5,59.0) (6,58.33) (7,60.71) (8,61.38) (9,48.3)};
\addlegendentry{Prefix-Z}
\addplot coordinates {(1,100.0) (2,91.67) (3,77.56) (4,64.83) (5,64.0) (6,59.22) (7,54.43) (8,57.29) (9,50.0)};
\addlegendentry{Prefix-R}
\addplot coordinates {(1,100.00) (2,96.67) (3,87.22) (4,65.42) (5,66.11) (6,60.00) (7,58.33) (8,60.54) (9,53.52)};
\addlegendentry{Prefix-A}
\end{axis}
\end{tikzpicture}
\caption{Ablation study of the impact of parallel adapter for long task on MMFi.}
\label{fig_ablation}
\vspace{-0.38cm} 
\end{figure}
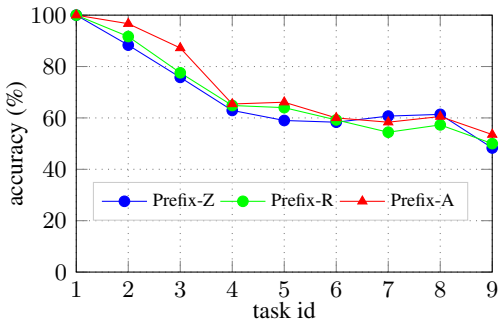

\subsubsection{Effect of Parallel Adapter}
Figure \ref{fig_ablation} shows that the parallel adapter initialization (Prefix-A) significantly outperforms zero (Prefix-Z) and random (Prefix-R) initializations in ConSense.
It achieves an average accuracy of 71.97\%.
In contrast, Prefix-R attains 69.00\% accuracy, while Prefix-Z achieves 68.30\%.
This highlights the effectiveness of the parallel adapter in enhancing prefix handling and overall model performance.

\vspace{-0.30cm} 
\section{Conclusion}

We propose ConSense, a lightweight and fast-adapted exemplar-free class incremental learning framework for WiFi-based HAR.
By leveraging the transformer architecture, ConSense effectively handles the challenges of continual learning in dynamic environments. The framework's key innovations, including dynamic model expansion on MHSA and selective retraining on MLP, enable fast training by focusing on integrating new information while preserving previously acquired knowledge.
Comparative tests on three datasets show that ConSense improves average accuracy by over 10\% across all tasks and maintains a forgetting rate below 20\%, outperforming existing methods.
Moreover, it reduces model parameters significantly, making it ideal for resource-constrained environments.
Ablation studies highlight the effectiveness of two strategies and parallel adapters in enhancing stability and accuracy.
Future efforts will focus on real-world applications and further optimization for edge deployments.

\section{Acknowledgments}
This work is supported by Natural Science Foundation of Jiangsu Province, China (Grant No. BK20230477 and BK20230482), National Science Foundation of China (NSFC No. 62302328 and No. 62106167), and the Priority Academic Program Development of Jiangsu Higher Education Institutions, Suzhou Frontier Science and Technology Program (Project SYG202310). 
\bibliography{aaai25}

\end{document}